\documentclass[runningheads]{llncs}
\usepackage[T1]{fontenc}
\usepackage{graphicx}
\usepackage{amssymb}
\usepackage{hyperref}

\usepackage[dvipsnames]{xcolor}
\usepackage{url}
\usepackage{siunitx}
\usepackage{amsmath}
\usepackage{eurosym}
\usepackage[normalem]{ulem}

\begin{document}
\title{Motion Analysis of Upper Limb and Hand in a Haptic Rotation Task}
%
%
\author{Kathrin Krieger\inst{1,2}\orcidID{0000-0003-2837-9887} \and
Yuri De Pra\inst{3}\orcidID{0000-0002-0876-4828} \and
Helge Ritter\inst{2}\orcidID{0000-0003-1703-1906}
\and
Alexandra Moringen\inst{2,4}\orcidID{0000-0002-2605-1020} 
}
\authorrunning{K. Krieger et al.}
%
\institute{Leibniz-Institut für Analytische Wissenschaften-ISAS-e.V., 44139 Dortmund, Germany, \email{kathrin.krieger@isas.de}\and
Neuroinformatics, Technical Faculty, CITEC, Bielefeld University, 33619 Bielefeld, Germany \email{\{kkrieger,helge\}@techfak.uni-bielefeld.de}\and
Research Centre ENEA CASACCIA, Via Anguillarese 301 00123, Italy\\
\email{ yuri.depra@enea.it} \and
Data Science Institute, University of Greifswald, Germany\\
\email{alexandra.moringe@uni-greifswald.de}}
\maketitle              
\begin{abstract}

Humans seem to have a bias to overshoot when rotating a rotary knob blindfolded around a specified target angle (i.e. during haptic rotation). Whereas some influence factors that strengthen or weaken such an effect are already known, the underlying reasons for the overshoot are still unknown. This work approaches the topic of haptic rotations by analyzing a detailed recording of the movement.
We propose an experimental framework and an approach to investigate which upper limb and hand joint movements contribute significantly to a haptic rotation task and to the angle overshoot based on the acquired data.
With stepwise regression with backward elimination, we analyze a rotation around \ang{90} counterclockwise with two fingers under different grasping orientations.  
Our results showed that the wrist joint, the sideways finger movement in the proximal joints, and the distal finger joints contributed significantly to overshooting.
This suggests that two phenomena are behind the overshooting:
1) The significant contribution of the wrist joint indicates a bias of a hand-centered egocentric reference frame. 
2) Significant contribution of the finger joints indicates a rolling of the fingertips over the rotary knob surface and, thus, a change of contact point for which probably the human does not compensate.

\end{abstract}
\section{Introduction}

In everyday life, people use rotary knobs in a wide variety of scenarios, such as operating the washing machine, settings in the car or music production equipment.
Manipulating rotary knobs for stroke patients with slightly impaired hand function is one of the functions they most want to recover~\cite{Lambercy2007AHK}.
Nevertheless, we know surprisingly little also about how healthy people perform haptic rotations. 

This publication investigates which movements are relevant for operating rotary knobs. To focus only on hand movements, we will investigate motion during \emph{haptic rotations}, i.e. rotations that are performed without vision and only by touch~\cite{Krieger2018IOS}.
So far, it has been found that if people aim at performing a haptic rotation around a target angle, they have a bias to overshoot~\cite{Krieger2018IOS,Krieger2019NOF,De2022EOR}.
However, the amount of overshooting differed across different experiments. 
Several studies investigated the participants' performance aiming at a \ang{90} counterclockwise rotation.
By measuring the signed error (i.e. the mean difference between the target and the encoded positions), a pilot study revealed that participants rotated on average \ang{62} too far.
Experiments with a standardized body and grasping posture resulted in an average overshooting of \ang{13}~\cite{Krieger2018IOS}, \ang{25}~\cite{Krieger2019NOF}, and \ang{38}~\cite{Krieger2019NOF}.
In a similar experiment, where participants were tasked to rotate \ang{45} and \ang{90} clockwise and counterclockwise, they rotated on average \ang{13} to \ang{15} too far, dependent on the experimental condition~\cite{De2022EOR}.
It is interesting to note that the overshoot was found in different studies imposing different upper limb constraints on the participants: in ~\cite{Krieger2018IOS,Krieger2019NOF} participants were grasping a knob placed parallel to the forearm (see Appendix, Section~\ref{app:markers}) while in ~\cite{De2022EOR} participants were grasping a knob placed perpendicularly to their forearm (i.e. in front of them).
Concerning the repeatability of such an overshoot, participants showed consistent movements across trials reporting a mean variable error (i.e. the standard deviation of the signed errors across trials) about \ang{10}~\cite{De2022EOR} and \ang{15}~\cite{Krieger2019NOF}.

Some factors that influence the amount of overshooting or repeatability are already known: it was shown that the rotary knob's shape features under the fingertips influence the mean signed error.
A flat shape leads to the best, an edge to medium, and a round shape to the worst result, i.e. more overshooting~\cite{Krieger2018IOS}. 
Moreover, grasping the rotary knob with more fingers during the haptic rotation leads to lower mean signed errors, thus, less overshooting~\cite{Krieger2019NOF}.
The distance covered and the direction did not show significant differences concerning both the signed and the variable error~\cite{De2022EOR}.

A predefined grasping orientation with respect to the rotational axis of the rotary knob does not impact the accuracy. 
However, it affects the precision, which expresses how similar the results are to each other when the experiment is repeated several times under identical conditions~\cite{Krieger2019NOF}. Without any experimental constraint, instead, the direction of the rotation significantly affected the initial grasping position of the fingers~\cite{De2022EOR}.
Although the reason for the bias towards overshooting is still unknown, we can analyze the movements involved in the rotation task to infer which part of the upper limb and the hand contributes more, and consequently influences more the hand proprioception during haptic rotation.

Movement strategies in rotations were analyzed in an experiment that evaluated a rotation task around at least \ang{270} with multiple re-grasps and visual feedback~\cite{Gurari2007HPI}.
The impact of three rotary knob sizes and two grasp types was investigated; the grasp types were a horizontal grasp, where the hand palm was parallel to the rotary knob's rotation axis, and a vertical grasp, where the back of the hand palm was perpendicular.
To analyze the movement, the authors identified four movement strategies based on video recordings: finger movement, elbow pronation/supination (rotation around the forearm's axis), wrist adduction/abduction, and wrist circumduction (combination of wrist adduction/abduction and flexion/extension).
For each trial, the most used of these strategies was annotated.
The results were a more proximal, i.e. closer to the body's center,  motion for the parallel perpendicular grasp, and for bigger rotary knobs.
The experiment revealed that movements like finger movements, wrist flexion/extension, wrist adduction/abduction, and elbow pronation/supination, contributed to rotary knob rotations in varying amounts and contingent upon the circumstances.
However, the authors noted that the rotations were often a combination of several movement strategies~\cite{Gurari2007HPI}. Building upon these results valuable in robotics, rehabilitation, or interaction scenarios, we further analyze movements during rotation tasks based on individual joint angle trajectories instead of annotated categories,  to gain insights about the contribution of each joint to the resulting rotation angle and to the resulting overshoot.

We analyze the motion data recorded in an experiment previously described in~\cite{Krieger2019NOF}, where study participants performed haptic rotations around \ang{90}.
Humans usually adapt their grasping orientation to the target angle~\cite{De2022EOR}. However, some grasping orientations with the same target angle lead to less precision~\cite{Krieger2019NOF}.
Since the grasping orientation impacts the haptic rotation process, it might influence the joint movements. 
Therefore, this work evaluates the contribution of the joint movements to the resulting rotation angle separately for different grasping orientations. 
Due to scope restrictions, we have included all technical figures referenced in this paper in the Appendix. Appendix is a separate file that can be found under the link: \url{https://nc.isas.de/s/tsRaf9jSiErtAk5} with Passwort: Eurohaptics\_2024

\section{Methods}
The experimental settings were described in detail in previous work~\cite{Krieger2019NOF}. In this paper, we will briefly revise the details relevant to the presented evaluation, while we want to refer the reader to~\cite{Krieger2019NOF} for a detailed description of the "mother" experiment.

\subsection{Participants}
Twenty sighted people (11 males, 9 females) with an age range of 19 to 34 years and without haptic impairments participated in the study.
However, due to motion data recording issues, only data from 18 participants were included in this current evaluation.
They signed an informed consent, and the ethics committee of Bielefeld University approved the experimental protocol.
The whole experimental session took one hour and participants received 6\,\euro to compensate for their time.

\subsection{Experimental Conditions}

\begin{figure}[h] 
	\centering
	\includegraphics[width=0.38\linewidth]{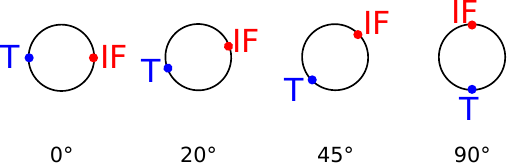}
	\caption{Grasping orientations from the top view. \textit{T} denotes the thumb's position and \textit{IF} the index finger's position on the rotary knob. Participants were standing, in relation to the figure's displayed top view, below the figure.}
	\label{fig:grip_startoris}
\end{figure}

This publication evaluates the results of five experimental conditions.
Since this experiment had a within-subjects design, all participants executed all experimental conditions. 
In all five experimental conditions, participants grasped the rotary knob with two fingers, namely the thumb ($T$) and the index finger ($IF$), from the top.
The difference between the five experimental conditions was the grasping orientation, meaning the position of the fingers with respect to the rotary knob.
In one experimental condition, participants were allowed to grasp freely, and in the other conditions, they had to grasp at the angles \ang{0}, \ang{20}, \ang{45}, and \ang{90}, as shown in Fig.~\ref{fig:grip_startoris}.
Each experimental condition was repeated eight times for each participant.

\subsection{Experimental Procedure}
First, participants signed the informed consent, and it was ensured that they had a clear understanding of \ang{90} rotation.
Reflective Vicon markers were placed on their body for motion tracking (see Appendix, Section~\ref{app:study}).
To allow participants to maintain a standardized body posture (see  Appendix, Section~\ref{app:posture}), a height-adjustable platform was adopted.
Participants were blindfolded, and the experiment started.

Each trial had the following procedure:
Participants took the standardized body posture; they placed their right hand in a starting position, as required by the motion tracking software.
The experimenter started data recording and locked the rotary knob mechanically to prevent participants from rotating it accidentally during grasping.
After the experimenter cued them to start, they grasped the round rotary knob from the top with the right hand.
The thumb and index finger had to touch the rotary knob with approximately half of the fingertip.
To enable better motion tracking, they had to leave the other fingers straight.
For the experimental conditions, which required a specific grasping orientation, the blindfolded participants had to haptically search for two elevations on the rotary knob and place their fingers exactly above.
Since two elevations create an ambiguous situation, participants always had to position their index finger either on the top, the top-left, or the left side.
The experimenter checked the grip and asked for correction if necessary.
The rotary knob was unlocked, and participants performed the \ang{90} counterclockwise rotation.
Afterward, they released the rotary knob carefully and placed their hand back in the starting position.

\subsection{Data Recording}
To investigate the relationship between joint movement and the resulting rotary knob rotation, two different data streams were captured and synchronized during the trial: the Vicon~\cite{ViconNexus25} markers trajectories, and the corresponding rotation angle of the knob. For control purposes only, Basler camera in top view was used to record the experiment on video.
To ensure synchronized recordings, the software \textit{multiple start synchronizer} (MSS) developed by~\cite{Maycock2010AMI} was used.
The knob rotation data stream was recorded by the Twister apparatus~\cite{Moringen2017HIT,Krieger2018IOS} and contained the rotary knob's orientation.
The marker trajectories were recorded by the Vicon system. We ensured the synchronization of both modalities by a common recording trigger. 
Moreover, to derive actual body joint angles from the recorded motion data, additional measures were taken from each participant once: an image of the hand on graph paper, and the measurement of the shoulder width.
\paragraph{\textbf{Rotary Knob Control and Recording with TWISTER}}
The details of the Twister apparatus used to record the rotary knob orientation are reported in~\cite{Moringen2017HIT,Krieger2018IOS}.
It has a table on which participants can place their hands, which was relevant to creating a standardized starting position.
Furthermore, it has an integrated shaft on which rotary knobs can be mounted.
For this data recording, a round rotary knob was chosen with a diameter of 40\,mm and a height of 70\,mm.
The rotary knob can be rotated or \textit{twisted} with almost zero resistance by participants or by an integrated motor placed under the table in a hidden cupboard.
The latter was used to orient the rotary knob with the attached elevations corresponding to the experimental condition.
Moreover, the Twister measures the shaft's, and therefore rotary knob's, absolute orientation with a resolution of \ang{0.022}.
It also detects whether participants touch the rotary knob.
The touch status and rotary knob orientations are recorded at 2\,kHz and sent via USB to software running on a computer.
Thus for each experimental trial, we recorded a rotary knob orientation trajectory $\boldsymbol\theta_{\tau}$.
\paragraph{\textbf{Motion Data}}
To track the motion of the upper limb and hand, reflective markers were attached to the body.
The marker placement is illustrated in Appendix, Section~\ref{app:markers}. 
The markers were recorded with 200\,Hz by 16 low-latency, high-speed, infrared motion MX3+ series Vicon cameras~\cite{ViconHardwarePDF}, which were placed around the participant's standing position. 
Based on the camera recordings, the corresponding software Vicon Nexus 2.5~\cite{ViconNexus25} calculated and stored the 3D positions of all markers in space.

\subsection{Joint Angle Computation}
The recorded 3D marker position trajectories were used to estimate the joint angles.
Since the joint angle computations of the hand and the upper limb differ, they are described separately.

\subsubsection{Hand}
\begin{figure}[h]
	\centering
	\includegraphics[width=0.6\linewidth]{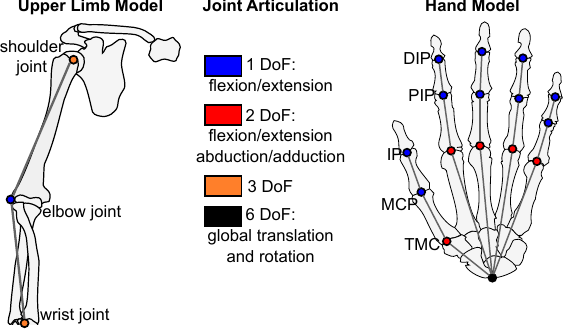}
	\caption{Schematic representation of the upper limb and hand model used in this work to estimate joint angles based on motion data. The upper limb model was created based on the example in~\cite{Yahya2019MCS}. The hand model is part of the automatic motion tracking software~\cite{Maycock2015FAO} and was created by~\cite{Schroder2014RTH}. 
 }
	\label{fig:models}
\end{figure}

To accurately estimate the joint angles of the hand from typically incomplete motion data provided by VICON, we used the motion tracking software described in~\cite{Maycock2015FAO}. 
The software uses a model-based inverse kinematics approach to label the trajectories, fill data gaps, and calculate finger joint angles~\cite{Maycock2015FAO}.
It builds mainly upon the kinematic hand model created by\cite{Schroder2014RTH} and schematically shown in Fig.~\ref{fig:models}, right. 
The hand model consists of 16 joints with 26 DoF.
The first joint is the origin with 6 DoF for global translation and rotation.
Each finger has three joints, of which the first one has 2 DoF, and the other two joints have 1 DoF~\cite{Schroder2014RTH}.

For each participant, we created a customized kinematic hand model.
The motion tracking software version allowed several adaptations.
Based on the image of the hand on the graph paper, the kinematic hand model's fingers were appropriately scaled, and the exact marker placement on the model was specified. 
For each trial, we loaded the participants' hand model and marker placements as well as the 3D position trajectories of all tracked markers. 
The motion tracking software returned the 20 joint angle trajectories and six rotation and translation trajectories for the total 26 DoF of the hand model~\cite{Maycock2015FAO}.

Importantly, for the current evaluation, only the joint angle trajectories ($\theta$) of the thumb ($t$) and index finger ($i$) are relevant.
The thumb had the three joints $TMC$, $MCP$, $IP$, while the index finger had the joints $MCP$, $PIP$, $DIP$, see Fig.~\ref{fig:models}, right.
Each joint had at least one DoF, which is a flexion vs. extension articulation ($f/e$).
The first joint of each finger, i.e., the TMC of the thumb and the MCP of the index finger, had an additional DoF, namely abduction vs. adduction ($ab/ad$).
An illustration of the $f/e$ and $ab/ad$ movement directions is in the Appendix, Section~\ref{app:movement-direction}.
Summarizing, for each trial, we had eight finger joint angle trajectories\footnote{With bold small letters, we will denote one-dimensional trajectories or time-series corresponding to one particular angle. With a bold capital letter, we will denote a multimodal time-series corresponding to multiple joint trajectories recorded during a particular time interval. For improved readability, we will drop the time indices. }: 
\begin{equation*} 
\boldsymbol{\theta^{ab/ad}_{t,TMC}, \theta^{f/e}_{t,TMC}, \theta^{f/e}_{t,MCP}, \theta^{f/e}_{t,IP}, \theta^{ab/ad}_{i,MCP}, \theta^{f/e}_{i,MCP}, \theta^{f/e}_{i,PIP}, \theta^{f/e}_{i,DIP}}
\end{equation*}

\subsubsection{Upper Limb Joints}
The upper limb joint angles were derived from the 3D marker position trajectories in two steps.
First, two rigid models were tracked with the motion tracking software~\cite{Maycock2015FAO}, which means that the 3d position trajectories of the corresponding model Vicon markers were labeled.
We used one rigid model for the shoulder and one for the elbow, see  Appendix, Section~\ref{app:markers}. 
Thereafter, a Python script estimated the joint angles of the upper body.
The decision on which joints to calculate was based on an upper limb model with three joints and, in total, 7 DoF presented in~\cite{Yahya2019MCS}.
Deviating from this, however, we assigned the rotation joint between the elbow and wrist to the wrist. 
The resulting upper limb model is shown in Fig.~\ref{fig:models}, left. The outcome of the computation was 3 DoF for the shoulder ($s$), 1 DoF for the elbow ($e$), and 3 DoF for the wrist joint ($w$), see Fig.~\ref{fig:models}, left.
Based on the used reference frames, the 3 DoF articulations are towards the $x$, $y$ or $z$ orientation.
This results in seven joint angle trajectories for each trial: $\boldsymbol{\theta^x_s, \theta^y_s, \theta^z_s, \theta_e, \theta^x_w, \theta^y_w, \theta^z_w}.$ 
The estimation of the joint angles is described and schematically illustrated in Appendix Section~\ref{app:markers}.

\subsection{Data Preprocessing}

In this paragraph, we focus on two steps of preprocessing of the recorded data: extraction of the relevant segment containing the rotation data only,  followed by the calculation of the criterion and the predictor variable values based on the extracted segment. 
For each trial, we get eight trajectories for the finger joints, seven trajectories for the upper limb joints, and a trajectory with the rotary knob orientations, denoted by $\boldsymbol{\Theta}$: 
\begin{equation*}
\begin{split}
 \boldsymbol{\Theta} = 
 &( \boldsymbol{\theta_{\tau}}\\
& \boldsymbol{\theta^{ab/ad}_{t,TMC}, \theta^{f/e}_{t,TMC}, \theta^{f/e}_{t,MCP}, \theta^{f/e}_{t,IP}, \theta^{ab/ad}_{i,MCP}, \theta^{f/e}_{i,MCP}, \theta^{f/e}_{i,PIP}, \theta^{f/e}_{i,DIP},} \\
& \boldsymbol{\theta^x_s, \theta^y_s, \theta^z_s, \theta_e, \theta^x_w, \theta^y_w, \theta^z_w,} ).
\end{split}
\end{equation*}
The trajectories of one trial are illustrated in  Appendix, Section~\ref{app:trajecoty}.

A complete trial consisted of several movement segments, one from the starting position to the rotary knob, the grasping, the rotation, and the releasing of the rotary knob.
Apart from the relevant segment, i.e. the rotation of the knob, the other parts of the recorded movement are not relevant to this work and had to be discarded. In the first preprocessing step, we extract the trajectory segments that contain just the haptic rotation movements  $\boldsymbol{\Theta_{start:end}}$ based on the estimated rotation time span (abbreviated by [start:end])\footnote{This estimation has been performed with the help of an algorithm, checked visually, and manually corrected, if necessary. }.

Both the criterion variable and the predictor variables are calculated in the second preprocessing step based on the extracted segment $\boldsymbol{\Theta_{start:end}}$:
 \begin{equation*} 
 \boldsymbol{\delta}=\max(\boldsymbol{\Theta_{start:end}})-\min(\boldsymbol{\Theta_{start:end}}), 
 \end{equation*} 
where maximum and minimum are calculated over time for each joint angle dimension, respectively,  and the subtraction is performed on the resulting vectors. For  
one trial, we denote the resulting range values as follows: 
\begin{equation*} 
\begin{split} 
\boldsymbol{\delta} = 
  &(\delta_{\tau}, \\
  &\delta^x_s,\delta^y_s, \delta^z_s, \delta_e, \delta^x_w, \delta^y_w, \delta^z_w, \\
  & \delta^{ab/ad}_{t,TMC}, \delta^{f/e}_{t,TMC}, \delta^{f/e}_{t,MCP}, \delta^{f/e}_{t,IP}, \delta^{ab/ad}_{i,MCP}, \delta^{f/e}_{i,MCP}, \delta^{f/e}_{i,PIP}, \delta^{f/e}_{i,DIP}),
\end{split}
\end{equation*} 
where $\delta_{\tau}$ denotes the resulting rotation performed by the study participant. 

For the purpose of further regression analysis, we split all data in two sets, one with data of study participants' joint angle ranges $\boldsymbol{\Delta_j}$ (the observed values of the predictor variable), and one with the resulting rotary knob  angle range $\boldsymbol{\Delta_\tau}$ (the observed values of the criterion variable). For all trials $i \in T$:
\begin{equation}
\{ \boldsymbol{\delta} =(\boldsymbol{\delta_j}, \boldsymbol{\delta_\tau}) \in \boldsymbol{\Delta_j} \times \boldsymbol{\Delta_\tau} | \boldsymbol{\delta_j} \in \mathbb{R}^{15}, \boldsymbol{\delta_\tau} \in \mathbb{R} \}.
\end{equation}
  We drop the trial index in this text and write $\boldsymbol{\delta}$ instead of $\boldsymbol{\delta^{i}}$ to simplify the notation.

\section{Results} 
\paragraph{\textbf{Relationship between joints and overshooting}}
To examine which joints contribute significantly to overshooting, we performed a regression analysis.
The criterion variable was accuracy, i.e. $\delta_{\tau}-\ang{90}$.
The predictor variables are the joint angle ranges $\delta_j$. 
Because several joints have large ranges and others comparatively small ones, using the raw values may  result in overlooking the influence of the joints with a small range. We performed a min-max normalization to get all values into the range $[0,1]$ per participant per joint, resulting in $\Delta_{j,norm}$.
To find which joints contribute significantly to the overshooting, a stepwise regression with backward elimination was executed.
It had the following procedure:
First, all parameters were included, and the model was fitted.
Second, the parameter with the highest non-significant p-value was eliminated.
Then, the model was fitted again with the remaining parameters.
The procedure was repeated until only parameters with significant p-values remained.
This procedure was executed with a Python script using the library statsmodels.api and the linear model ordinary least square regression.

\begin{table*}[ht]
\caption{Regression coefficients of the regression models' last step. More detailed results in Appendix, Section~\ref{app:results-detailed}. \textit{Left:} Regression to investigate contributions to overshooting. The significant predictor variables are shown. A green number indicates a contribution to more rotation, i.e. to overshooting, while a red one indicates a contribution to less rotation, i.e. compensation of the overshooting. \textit{Right:} Regression to investigate the joints' contribution to the rotary knob rotation for all grasping orientations separately, \ang{0}, \ang{20}, \ang{45}, \ang{90} and free. A green number indicates a positive and red a negative relationship between the used range of motion and the rotary knob rotation.}
\label{table:results}
\centering
\scriptsize
 \begin{tabular}{lccccc} 
 \hline
 predictor& regression coef &  &  &  & \\
  \hline
  $\delta_{e,norm}$ &\color{red}{-13.62}&&&&\\
  $\delta^x_{w,norm}$ &\color{Green}{~30.61}&&&&\\
  $\delta^y_{w,norm}$ &\color{Green}{~21.45}&&&&\\
  $\delta^z_{w,norm}$ &\color{Green}{~20.13}&&&&\\
  $\delta^{ab/ad}_{t,TMC,norm}$ &\color{red}{-17.46}&&&&\\
  $\delta^{f/e}_{t,TMC,norm}$ &\color{Green}{~24.05}&&&&\\
  $\delta^{f/e}_{t,MCP,norm}$ &\color{red}{-25.96}&&&&\\
  $\delta^{f/e}_{t,IP,norm}$ &\color{Green}{~17.74}&&&&\\
  $\delta^{ab/ad}_{i,MCP,norm}$ &\color{Green}{~~8.55}&&&&\\
  $\delta^{f/e}_{i,DIP,norm}$ &\color{Green}{~34.67}&&&&\\
  \hline
 \end{tabular}
 \quad
 \begin{tabular}{lccccc} 
 \hline
predictor & \ang{0} & \ang{20} & \ang{45} & \ang{90} & free\\
  \hline
  const &\color{Green}{~28.36}&\color{Green}{~62.22}&\color{Green}{~67.52}&\color{Green}{~70.50}&\color{Green}{~66.18}\\
  $\delta^x_s$ & &\color{red}{~-7.03}&\color{red}{~-3.86}&\color{Green}{~~3.63}&\color{red}{~-6.03}\\
  $\delta^y_s$ &\color{Green}{~63.32}&\color{Green}{~40.89}&&&\\
  $\delta^z_s$ & &\color{Green}{~10.22}&\color{Green}{~17.70}&\color{Green}{~~9.28}&\color{Green}{~36.50}\\
  $\delta_e$ &\color{red}{-25.40}&&&&\\
  $\delta^x_w$ &\color{Green}{~11.85}&&\color{Green}{~20.83}&&\\
  $\delta^y_w$ &\color{Green}{~46.40}&&&&\\
  $\delta^z_w$ & &\color{Green}{~52.90}&&&\color{Green}{~27.95}\\
  $\delta^{ab/ad}_{t,TMC}$ & &&&&\color{Green}{~28.42}\\
  $\delta^{f/e}_{t,TMC}$ &\color{Green}{~51.07}&&\color{Green}{~27.98}&\color{Green}{~26.52}&\\
  $\delta_{t,MCP}$ &\color{red}{-46.51}&&&&\\
  $\delta_{t,IP}$ & &&\color{Green}{~17.89}&&\\
  $\delta^{ab/ad}_{i,MCP}$ &\color{Green}{~67.29}&\color{Green}{~55.47}&\color{Green}{~23.87}&\color{Green}{~43.85}&\\
  $\delta^{f/e}_{i,MCP}$ &\color{Green}{~75.91}&\color{Green}{~37.60}&\color{Green}{~52.28}&&\color{Green}{~41.50}\\
  $\delta_{i,PIP}$ & &&&\color{Green}{~22.02}&\color{Green}{~26.80}\\
  $\delta_{i,DIP}$ & &&\color{Green}{~29.34}&\color{Green}{~40.23}&\color{Green}{~61.57}\\
    \hline

 \end{tabular}
\end{table*}

The results of the last step are shown in shown in Table~\ref{table:results}, left.
A positive regression coefficient indicates a contribution to more rotation, i.e. to overshooting, while a negative regression coefficient expresses less rotation, i.e. compensation of the overshooting.
The elbow had a negative relationship, which means that it compensates the overshooting.
For the wrist, all three rotation axes had a significant and positive impact, meaning that they strongly contributed to the overshooting.
For the thumb, the \textit{ab/ad}-movement of the TMC and the MCP movement contributed significantly negatively to the accuracy, i.e. reducing the overshooting.
On the contrary, the thumb's \textit{f/e}-movement of MCP and the IP had a positive significant impact on the accuracy, i.e. contributing to overshooting.
The index finger's \textit{ab/ad}-movement of the MCP and the DIP movement also contributed significantly positively, i.e. to overshooting.

\paragraph{\textbf{Contribution of the joints to the rotation}}
To additionally examine which joint angle ranges contributed significantly to the rotation, we performed an additional analysis.
Since \cite{Gurari2007HPI} showed that different grasping orientations might lead to different joint usage, we performed the analysis separately for each experimental condition. 
The criterion variable was $\Delta_{\tau}$.
The input or so-called predictor variables were all raw joint angle ranges $\Delta_{j}$.
Additionally, we added a constant to the predictor variables to compensate for additional body movements, like an upper body rotation around the vertical axis.
Similarly to the last section, a linear model, ordinary least square regression with stepwise backward elimination was executed.

Table~\ref{table:results}, right shows the results of all five regression analyses corresponding to the five investigated experimental conditions.
Due to the space limitation, just the significant regression coefficients are listed.
The table contains columns for all experimental conditions and rows for the parameters in $\Delta_j$.
However, coefficients are provided just for the parameters with a significant p-value in the last step of the regression.
For simpler visualization, positive relations are shown with green coefficients. 
They express that the more the joint movement was used, the higher the rotation angle.
Negative relations are indicated with red coefficients, meaning that the more the joint movement was used, the less the resulting rotation.
Thus, such movements can be considered as a way of compensation.


\section{Discussion}
\paragraph{\textbf{Relationship between joints and overshooting}}
The results have shown that there were three major contributions to overshooting. 
The first one was the wrist joint.
The underlying mechanisms for this are still not clear.
On the one hand, it is already known that the human's judgment of orientations is biased by an egocentric reference frame~\cite{Van2014HPO}.
Different egocentric reference frames can exist, and typically, the predominant egocentric reference frame is dependent on the task~\cite{Volcic2008AAE}.
Thus, it is likely that the hand is used as the main egocentric reference frame. 
In such a case, it might be that the movement of the hand itself is not taken enough into account.
However, the exact contribution to the overshooting of a possible existing egocentric reference frame in the hand should be investigated in further, more standardized experiments.

The second major contribution to the overshooting came from the finger joints.
For the thumb TMC joint, the flexion and extension movement contributed significantly, which describes a sideways movement, as shown in Fig~\ref{fig:models}.
For the index MCP joint, the abduction and adduction movement, which is also a sideways movement as shown in Fig.~\ref{fig:models}, provided a significant contribution.
An illustration of those movement directions is in Appendix, Section~\ref{app:movement-direction}.
For both involved fingers, a sideways movement in the first joint leads to overshooting.
Moreover, for both involved fingers, the flexion and extension movement of the last joint provided a significant contribution to overshooting.
Looking at the videos of the trials showed that a sideways finger movement of the first joint or a flexion movement of the last joint results in the finger changing its orientation towards the rotary knob.
A very simplified scenario is illustrated in the Appendix, Section~\ref{app:rolling}.
From a theoretical perspective, when bending these joints and having these degrees of freedom, the fingers must change their orientation with respect to the rotary knob.
This results in the fingertips  \textit{rolling} over the surface, i.e. changing the contact point.
By this contact point change, the rotary knob moves more than the finger itself has actually moved.
So it may be possible that humans are not aware, when blindfolded, that this \textit{rolling} mechanism leads to additional rotary knob movement, and thus, do not compensate for it.
\paragraph{\textbf{Contribution of the joints to the rotation}}
The results indicate that the \textbf{shoulder} joint contributed significantly to the rotary knob rotation in all experimental conditions. 
The movement direction of the shoulder joint differs depending on the grasping orientation. 
While in the condition with the \ang{0} grasping orientation, just one movement direction was used, the other conditions relied on a more complex shoulder movement.
Additionally, in the \ang{20} and \ang{45} grasping orientation conditions, one shoulder direction movement had a negative regression coefficient.
That indicates that those shoulder movements were used to reduce the rotation.
The \textbf{elbow} joint contributed significantly only in the condition with the grasping orientation \ang{0}.
It had a negative coefficient, meaning that the less the elbow was used, the more the rotary knob was rotated. 
However, the elbow joint angles should be taken with caution.
As it is visible in Fig.~\ref{fig:models}, the assumed elbow joint was not located in the middle of the actual elbow joint but instead on the outer side.
The reason was that, based on the tracked Vicon marker positions on the body, it was impossible to recover the middle of the actual elbow joint.
To estimate this elbow joint center, an additional Vicon marker on the other side of the elbow would have been required.
Unfortunately, due to the nature of the haptic rotation movement, the Vicon marker in this position would be occluded and not visible to the Vicon cameras most of the time.
Thus, using the outer side of the actual elbow for the model's elbow joint position was the best approximation.
Future research may improve the motion-tracking process and further investigate the elbow joint. 
The \textbf{wrist} joint contributed significantly in all experimental conditions except in the \ang{90} grasping orientation. 
This also seems reasonable since the rotary knob was grasped in this condition so that the hand extends the lower arm. 
To contribute to a counterclockwise rotation, the wrist joint would require movement in a direction that is very limited.
Therefore, it seems reasonable that in the \ang{90} grasping orientation condition, the rotary knob rotation was mainly realized by other joints. The \textbf{thumb} contributed significantly with different joints and movement directions in most cases, except in one experimental condition.
For the \textbf{index} finger, the results indicated that the flexion/extension movement of the MCP had a significant contribution in all grasping orientations.
For smaller grasping orientations, the abduction/adduction of the MCP joint had significant contributions, but the regression coefficient became smaller for larger grasping orientation angles.
Contrary to that, with larger grasping orientations, the significant coefficients of the index PIP and DIP become larger.
Thus, depending on the grasping orientation, either the MCP abduction/adduction or the DIP and PIP joints are used. 
\paragraph{\textbf{Future Work}}
This evaluation has revealed that two mechanisms may contribute to the overshooting, the wrist joint movement which may result from an egocentric reference frame bias in the hand and \textit{rolling} of the fingertips above the rotary knob surface.
Future work may investigate the impact of both phenomena more systematically by either restricting the hand movement or ensuring that no \textit{rolling} happens. 
Moreover, the not-yet-analyzed data from this experiment can be used to understand why humans overshoot less when using more fingers for the haptic rotation.
Until now, it was not known whether either their movement was restricted or whether they performed better due to more perceived information.
Finally, the data can also be used to evaluate other movement parameters like velocity or also other parts of the movement like the grasping or releasing process.

\begin{credits}
\subsubsection{\ackname} This work was supported by the DFG Center of Excellence EXC 277: Cognitive Interaction Technology (CITEC) and the German-Japan Collaborative Research Program on Computational Neuroscience (RI 621/9-1).
We want to thank Guillaume Walck for the massive amount of help with processing the motion data and extracting the joint angles.

\end{credits}
%
%
%
\bibliographystyle{splncs04}
\bibliography{bibfile}
\newpage
\appendix 
\section{Standardized body posture}
\label{app:posture}
\begin{figure}[h] 
	\centering
	\includegraphics[height=0.35\linewidth]{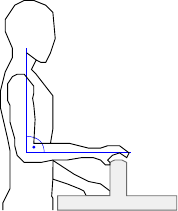}
    \includegraphics[height=0.35\linewidth]{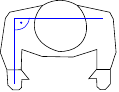}
	\caption{Standardized body posture from the side and top view.}
	\label{fig:posture}
\end{figure}

\newpage
\section{Finger movement directions}
\label{app:movement-direction}
\begin{figure}[ht!]
	\centering
    \includegraphics[width=0.49\linewidth]{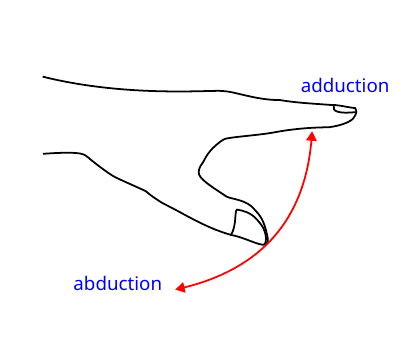}
    \includegraphics[width=0.49\linewidth]{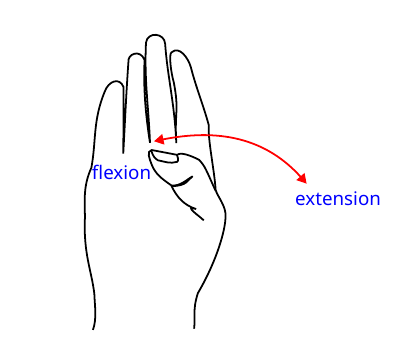}\\
    \includegraphics[width=0.49\linewidth]{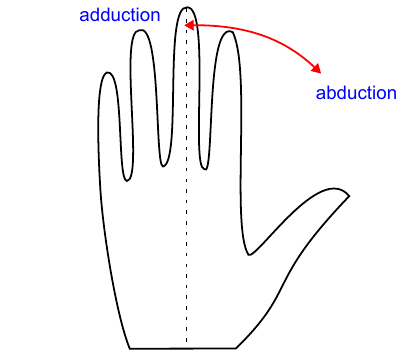}
    \includegraphics[width=0.49\linewidth]{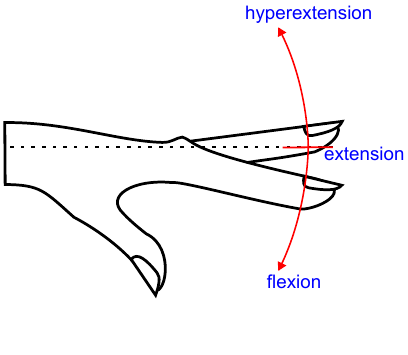}
	\caption{Schematic representation of finger joint movements. \textit{Top}: Movements of the thumb. \textit{Bottom}: Movements of the index finger. 
 }
	\label{fig:movement-direction}
\end{figure}

\newpage
\section{Illustration of the rolling mechanism}
\label{app:rolling}
\begin{figure}[h]
	\centering
    \includegraphics[width=\linewidth]{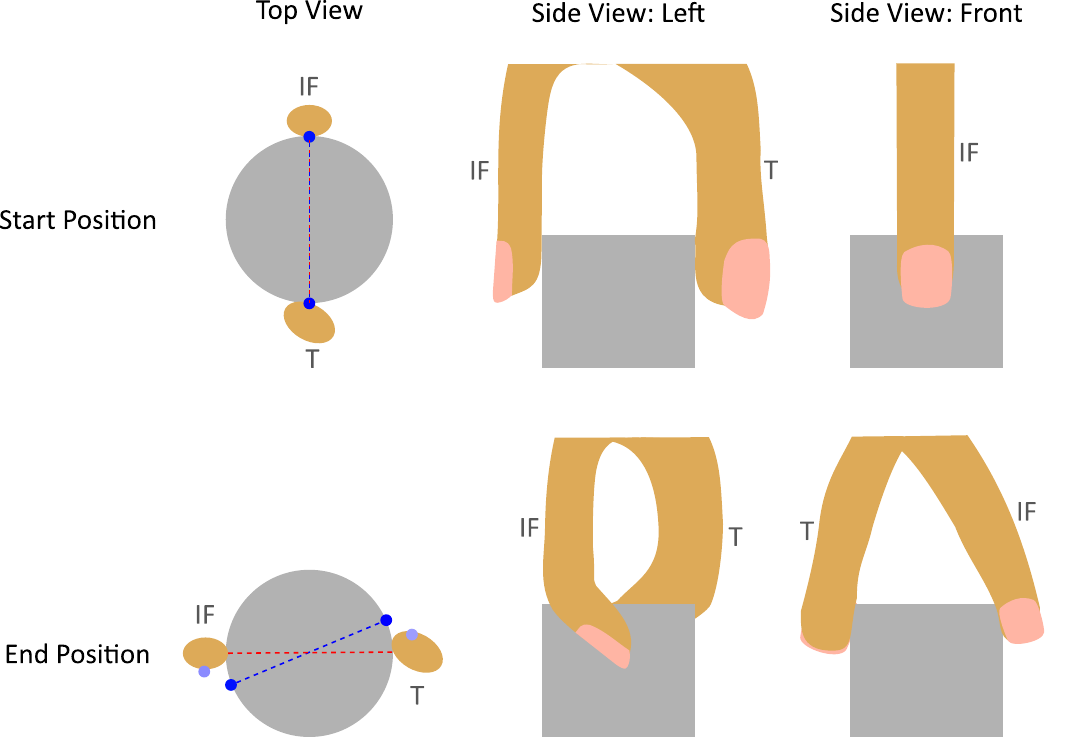}
	\caption{Illustration of the rolling mechanism. T denotes the thumb and IF the index finger. The column top view, row start position shows with the blue and red dot the initial contact point between the fingers and the rotary knob. The row end position shows the fingers after the \ang{90} rotation, while the initial contact points are still marked. It is visible, that since the fingers have \textit{rolled} over the rotary knobs surface, the new contact points are different. It also becomes apparent that in case the fingers perform a \ang{90} rotation, the rotary knob rotated slightly more. The side view columns are just for reference to show how the index finger's DIP is flexed, the MCP is abducted and the thumb's IP and TMC are flexed. 
 }

\end{figure}

\newpage
\section{Example trajectories for one trial}
\label{app:trajecoty}
\begin{figure}[h]
	\centering
	\includegraphics{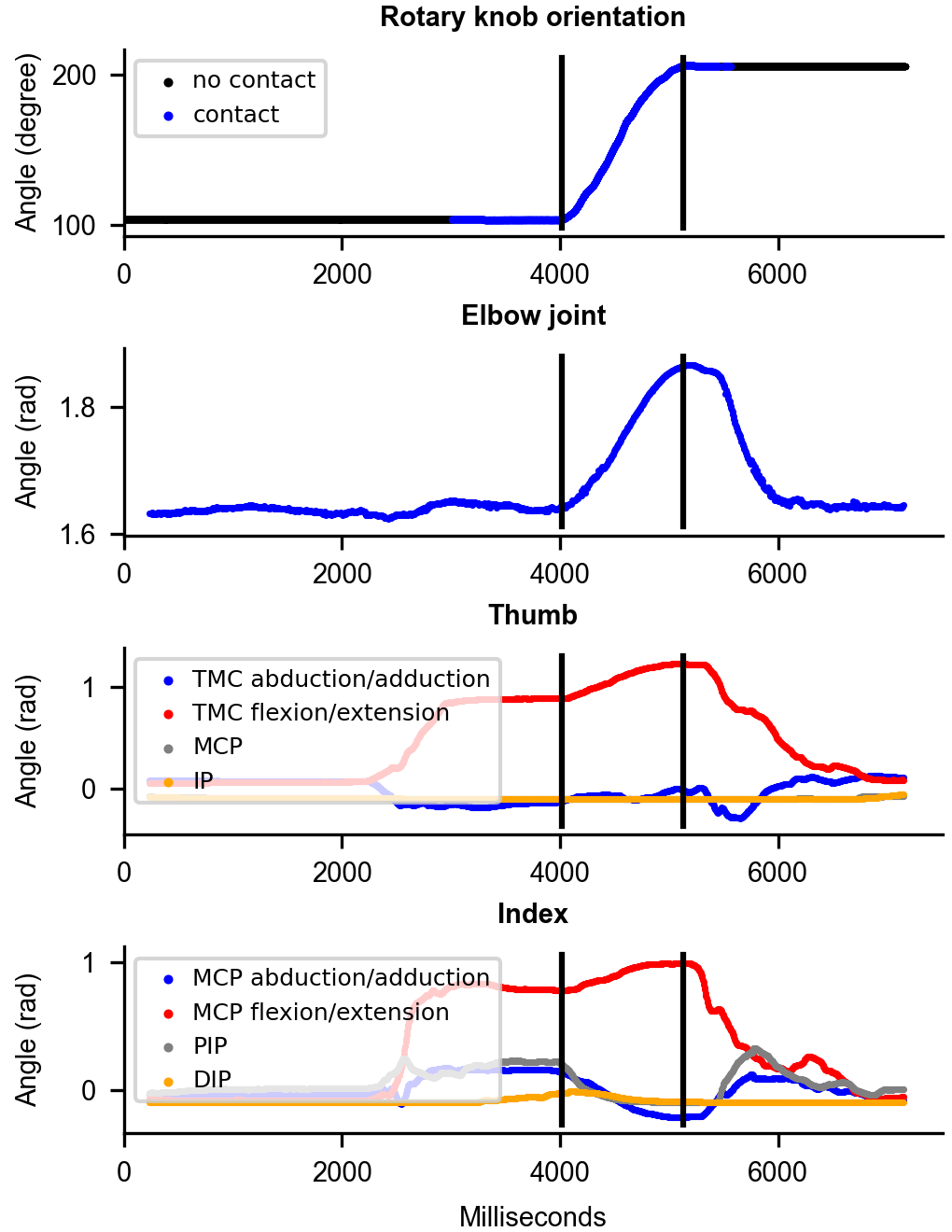}
	\caption{Example trajectories for one trial executed by participant 10. The $x$-axis interval between the vertical black lines corresponds to the rotation phase. The interval between two vertical black lines shows the estimated time span of the rotation movement. 
 }
	\label{fig:trajectory-example}
\end{figure}
\newpage
\section{Marker Placement on a Study Participant}
\label{app:study}

\begin{figure}[h]
 \centering
 \includegraphics{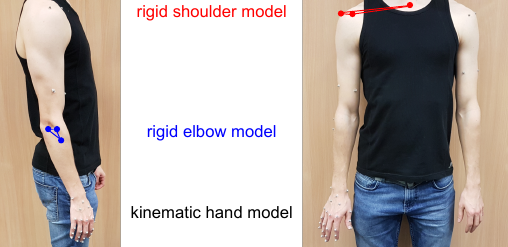}
 \caption{Vicon marker placement on upper limb and hand. Markers grouped together to models and used by motion tracking software are marked.}
 \label{fig:marker-placement}
\end{figure}
\newpage
\section{Marker Placement and Joint-angle estimation}

\label{app:markers}
\begin{figure}[h]
	\centering
	\includegraphics[width=0.6\linewidth]{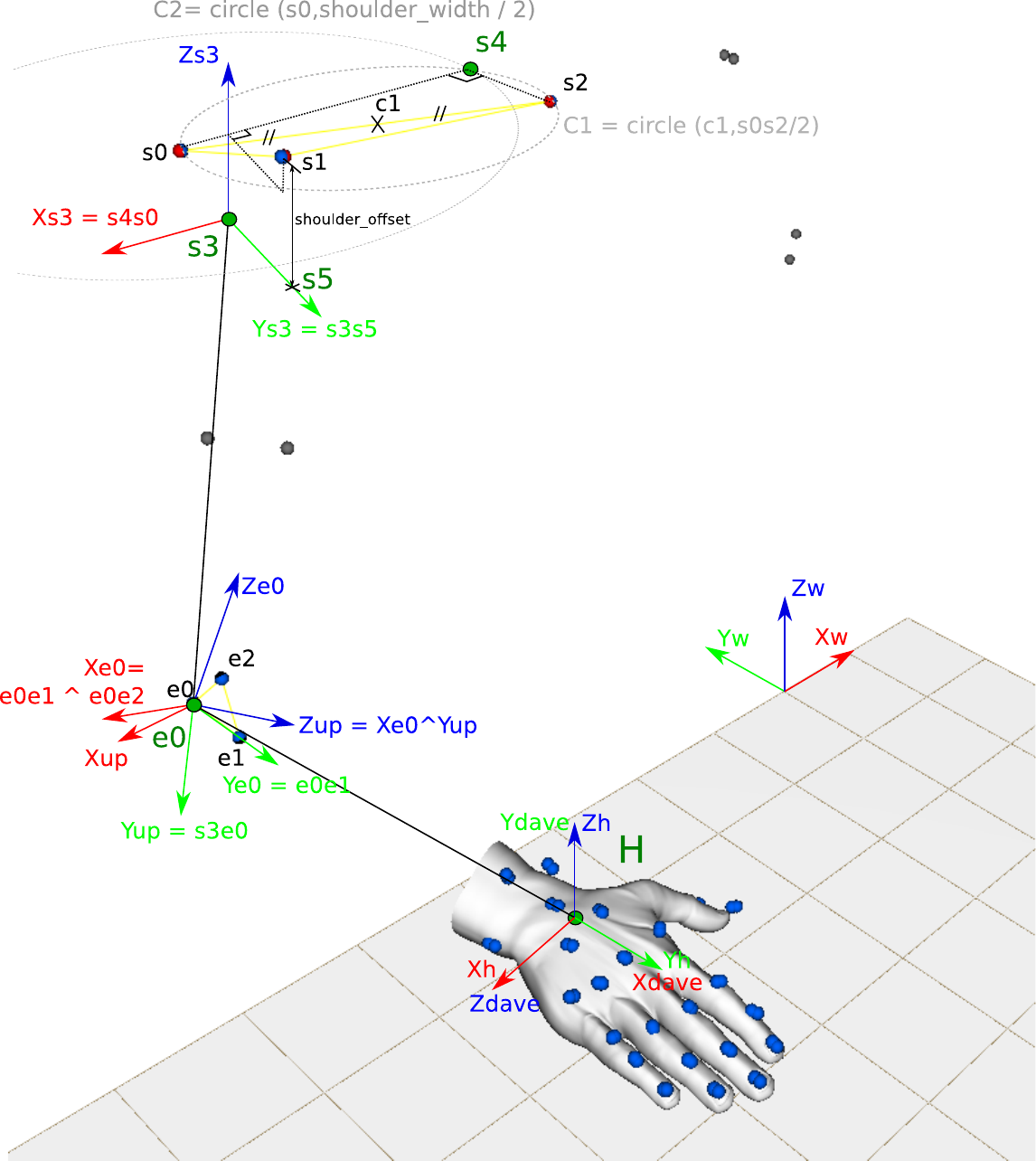}
	\caption{
 Schematic description of upper limb joint angle estimation.
 }
	\label{fig:upper-limb-joint-calculation}
\end{figure}
The geometrical constructions using the measured participants' shoulder width permit to recover the shoulder to shoulder axis (s4s0) from the 3 markers in the shoulder model, implicitly creating the shoulder frame $x$-axis. s2 denotes the marker in the middle of the neck. The shoulder $z$-axis is chosen upwards since the participants were asked to stand in a reference position, with torso straight, at the beginning of each trial. The $y$-axis is computed to get an orthonormal frame. The origin of the frame s3 is below the Scapular Acromion on the shoulder $x$-axis, at a constant offset. This frame is stored relatively to the frame of shoulder model at the beginning of each trial and recomputed from the tracked shoulder model at each time step.  
 The forearm frame is constructed based on the elbow model and its tracked frame. The $y$-axis is along the forearm (e0e1), the $x$-axis is the normal to the plane formed by the elbow markers (e0, e1, e2), and $z$-axis to form the orthonormal frame. The origin is at the non-moving marker of the elbow e0. Finally the upperarm frame at the elbow is built as follows. The y-axis is along the line between the shoulder center s3 and the forearm origin above the elbow joint e0. Such an origin is not ideal since the joint is effectively towards inside of the arm, but the true angles of the shoulder joint are not necessary, only relative motion is important in this study. The upperarm $z$-axis is perpendicular to the forearm $x$-axis and to upper arm $y$-axis, $x$-axis of the upperarm is computed to create an orthonormal frame. The hand pose is already extracted by the hand tracking software and defines the palm frame.

\newpage
\section{Detailed Regression Results}
\label{app:results-detailed}
\begin{table*}[ht]
\caption{Regression to investigate contributions to overshooting. The significant predictor variables are shown for the last, i.e. 6th step. $R^2=0.691$}
\label{table:results-overshooting-detailed}
\centering
 \begin{tabular}{lrrrrrr} 
 \hline
 predictor& coef &~~std err & $t$ & $p$ & [0.025&0.975]  \\
  \hline
  $\delta_{e,norm}$ &-13.62&5.41&-2.52&$0.012$&-24.24&-3.00\\
  $\delta^x_{w,norm}$ &30.61&5.20&5.90&$<0.001$&20.41&40.81\\
  $\delta^y_{w,norm}$ &21.45&6.08&3.53&$<0.001$&9.51&33.38\\
  $\delta^z_{w,norm}$ &20.13&4.44&4.54&$<0.001$&11.42&28.85\\
  $\delta^{ab/ad}_{t,TMC,norm}$ &-17.46&4.44&-3.93&$<0.001$&-26.19&-8.73\\
  $\delta^{f/e}_{t,TMC,norm}$ &24.05&5.54&4.34&$<0.001$&13.17&34.92\\
  $\delta^{f/e}_{t,MCP,norm}$ &-25.96&4.83&-5.37&$<0.001$&-35.45&-16.47\\
  $\delta^{f/e}_{t,IP,norm}$ &17.74&5.28&3.36&$0.001$&7.37&28.11\\
  $\delta^{ab/ad}_{i,MCP,norm}$ &8.55&4.30&1.99&$0.047$&0.11&16.99\\
  $\delta^{f/e}_{i,DIP,norm}$ &34.67&4.83&7.18&$<0.001$&25.19&44.15\\
  \hline
 \end{tabular}
\end{table*}

\begin{table*}[ht]
\caption{Regression model with the criterion $\Delta_{\tau}$ and all raw joint angle ranges $\Delta_{j}$ in the experimental condition where participants grasped the rotary knob at \ang{0}. The table shows the results of the regression models' last, i.e. the 8th step. $R^2=0.775$ }
\label{table:results-0deg-detailed}
\centering
 \begin{tabular}{lrrrrrr} 
 \hline
 predictor& coef &~~std err & $t$ & $p$ & [~0.025&~0.975~]  \\
  \hline
  const &28.36&7.44&3.81&$<0.001$&13.61&43.11\\
  $\delta^y_s$ &63.32&9.23&6.86&$<0.001$&45.03&81.60\\
  $\delta_{e}$ &-25.40&11.90&-2.14&$0.035$&-48.97&-1.83\\
  $\delta^x_{w}$ &11.85&4.12&2.88&$0.005$&3.69&20.01\\
  $\delta^y_{w}$ &46.40&9.59&4.84&$<0.001$&27.40&65.40\\
  $\delta^{f/e}_{t,TMC}$ &51.07&9.54&5.36&$<0.001$&32.17&69.97\\
  $\delta^{f/e}_{t,MCP}$ &-46.51&12.34&-3.77&$<0.001$&-70.96&-22.06\\
  $\delta^{ab/ad}_{i,MCP}$ &67.29&15.31&4.40&$<0.001$&36.95&97.63\\
  $\delta^{f/e}_{i,MCP}$ &75.91&16.27&4.67&$<0.001$&43.67&108.16\\
  \hline
 \end{tabular}
\end{table*}
\begin{table*}[ht]
\caption{Regression model with the criterion $\Delta_{\tau}$ and all raw joint angle ranges $\Delta_{j}$ in the experimental condition where participants grasped the rotary knob at \ang{20}. The table shows the results of the regression models' last, i.e. the 10th step. $R^2=0.799$}
\label{table:results-20deg-detailed}
\centering
 \begin{tabular}{lrrrrrr} 
 \hline
 predictor& coef &~~std err & $t$ & $p$ & [~0.025&~0.975~]  \\
  \hline
  const &62.22&5.95&10.46&$<0.001$&50.43&74.01\\
  $\delta^x_s$ &-7.03&1.83&-3.85&$<0.001$&-10.65&-3.40\\
  $\delta^y_s$ &40.89&11.22&3.64&$<0.001$&18.65&63.12\\
  $\delta^z_s$ &10.22&4.78&2.14&$0.035$&0.74&19.70\\
  $\delta^x_{w}$ &52.90&5.59&9.47&$<0.001$&41.83&63.97\\
  $\delta^{ab/ad}_{i,MCP}$ &55.47&13.40&4.14&$<0.001$&28.91&82.03\\
  $\delta^{f/e}_{i,MCP}$ &37.60&13.94&2.70&$<0.001$&9.98&65.23\\
  \hline
 \end{tabular}
\end{table*}

\begin{table*}[ht]
\caption{Regression model with the criterion $\Delta_{\tau}$ and all raw joint angle ranges $\Delta_{j}$ in the experimental condition where participants grasped the rotary knob at \ang{45}. The table shows the results of the regression models' last, i.e. the 8th step. $R^2=0.814$}
\label{table:results-45deg-detailed}
\centering
 \begin{tabular}{lrrrrrr} 
 \hline
 predictor& coef &~~std err & $t$ & $p$ & [~0.025&~0.975~]  \\
  \hline
  const &67.52&7.44&11.55&$<0.001$&55.95&79.09\\
  $\delta^x_s$ &-3.86&1.23&-3.12&$0.002$&-6.32&-1.41\\
  $\delta^z_s$ &17.70&2.37&7.46&$<0.001$&13.00&22.40\\
  $\delta^x_w$ &20.83&2.85&7.31&$<0.001$&15.19&26.47\\
  $\delta^{f/e}_{t,TMC}$ &27.98&6.54&4.28&$<0.001$&15.03&40.93\\
  $\delta^{f/e}_{t,IP}$ &17.89&6.51&2.75&$0.007$&5.00&30.79\\
  $\delta^{ab/ad}_{i,MCP}$ &23.87&11.25&2.12&$0.036$&1.60&46.15\\
  $\delta^{f/e}_{i,MCP}$ &52.28&12.60&4.15&$<0.001$&27.33&77.23\\
  $\delta^{f/e}_{i,DIP}$ &29.34&10.00&2.94&$0.004$&9.54&49.15\\
  \hline
 \end{tabular}
\end{table*}

\begin{table*}[ht]
\caption{Regression model with the criterion $\Delta_{\tau}$ and all raw joint angle ranges $\Delta_{j}$ in the experimental condition where participants grasped the rotary knob at \ang{90}. The table shows the results of the regression models' last, i.e. the 10th step. $R^2=0.639$}
\label{table:results-90deg-detailed}
\centering
 \begin{tabular}{lrrrrrr} 
 \hline
 predictor& coef &~~std err & $t$ & $p$ & [~0.025&~0.975~]  \\
  \hline
  const &70.50&5.00&14.10&$<0.001$&60.60&80.40\\
  $\delta^x_s$ &3.64&1.47&2.47&$0.015$&0.72&6.55\\
  $\delta^z_s$ &9.28&2.23&4.16&$<0.001$&4.87&13.70\\
  $\delta^{f/e}_{t,TMC}$ &26.52&7.75&3.42&$0.001$&11.18&41.86\\
  $\delta^{ab/ad}_{i,MCP}$ &43.85&11.63&3.77&$<0.001$&20.82&66.89\\
  $\delta^{f/e}_{i,PIP}$ &22.02&8.00&2.75&$0.007$&6.18&37.86\\
  $\delta^{f/e}_{i,DIP}$ &40.23&9.42&4.27&$<0.001$&21.57&58.89\\
  \hline
 \end{tabular}
\end{table*}

\begin{table*}[ht]
\caption{Regression model with the criterion $\Delta_{\tau}$ and all raw joint angle ranges $\Delta_{j}$ in the experimental condition where participants grasped the rotary knob freely. The table shows the results of the regression models' last, i.e. the 9th step. $R^2=0.617$}
\label{table:results-free-detailed}
\centering
 \begin{tabular}{lrrrrrr} 
 \hline
 predictor& coef &~~std err & $t$ & $p$ & [~0.025&~0.975~]  \\
  \hline
  const &66.18&5.40&12.25&$<0.001$&55.49&76.88\\
  $\delta^x_s$ &-6.03&1.60&-3.78&$<0.001$&-9.20&-2.87\\
  $\delta^z_s$ &36.50&6.18&5.91&$<0.001$&24.28&48.72\\
  $\delta^{ab/ad}_{t,TMC}$ &27.94&13.85&2.02&$0.046$&0.54&55.36\\
   $\delta^{f/e}_{t,TMC}$ &28.42&8.27&3.44&$0.001$&12.05&44.80\\
  $\delta^{f/e}_{i,MCP}$ &41.50&15.34&2.71&$0.008$&11.14&71.87\\
  $\delta^{f/e}_{i,MCP}$ &26.80&7.72&3.47&$0.001$&11.52&42.08\\
  $\delta^{f/e}_{i,DIP}$ &61.57&13.73&4.49&$<0.001$&34.41&88.74\\
  \hline
 \end{tabular}
\end{table*}

\end{document}